\documentclass[floatfix,aps,prd,amsmath,nofootinbib,preprintnumbers,onecolumn]{revtex4}
\pdfoutput=1

\usepackage{graphicx}
\usepackage{subfigure}
\usepackage{rotating}
\usepackage{float}

\def\({\left(}
\def\){\right)}
\def\[{\left[}
\def\]{\right]}

\def\e{\begin{equation}}
\def\q{\end{equation}}
\def\m{\begin{eqnarray}}
\def\n{\end{eqnarray}}

\begin{document}

\title{Signatures of Modified Dispersion Relation of Graviton in the Cosmic Microwave Background}

\author{Jun Li$^{1,2}$ \footnote{lijun@itp.ac.cn} and Qing-Guo Huang$^{1,2,3}$ \footnote{huangqg@itp.ac.cn}}
\affiliation{$^1$ CAS Key Laboratory of Theoretical Physics,\\ Institute of Theoretical Physics, \\Chinese Academy of Sciences, Beijing 100190, China\\
$^2$ School of Physical Sciences, \\University of Chinese Academy of Sciences,\\ No. 19A Yuquan Road, Beijing 100049, China\\
$^3$ Synergetic Innovation Center for Quantum Effects and Applications, Hunan Normal University, Changsha 410081, China}

\date{\today}

\begin{abstract}
The dispersion relation of graviton is a fundamental issue for fundamental physics about gravity. In this paper we investigate how the modified dispersion relation of graviton affects the cosmic microwave background (CMB) power spectra, in particular the B-mode polarization. Our results will be useful to test the dispersion relation of graviton at the energy scale around $10^{-29}$ eV.


\end{abstract}


\maketitle


\section{Introduction}

General Relativity is widely accepted as a fundamental theory to govern the dynamics of space-time itself. It governs the behaviors of black holes, the generation and propagation of gravitational waves, the formation of all structures and so on in the Universe.
However, the rapid developments of observations, in particular the evidence for the dark energy and dark matter, inspire some alternative theories to general relativity at some scales. Various attempts to modify gravity have been present in literatures. For example, see \cite{Clifton:2011jh,Capozziello:2011et,Nojiri:2010wj} for some recent reviews on modified gravity theories. Testing modified gravity theories in solar system, binary pulsar systems and cosmological scales has become one of the core tasks. In this paper we focus on the effects of modified gravity on the cosmic scales.

The cosmic microwave background (CMB) is a probe to physics in the early universe and the cosmological evolution that followed. Thomson scattering in the presence of primordial fluctuations affects not only the temperature but also its polarization. The CMB polarization which can be decomposed into E-mode and B-mode can be affected by both scalar and tensor perturbations. In particular, the B-mode component mainly comes from the tensor perturbation on very large scales and encodes the information of the gravitational waves \cite{Kamionkowski:1996zd,Cabella:2004mk,Radiationtransfer,Kosowsky:1998mb,Kamionkowski:2015yta}. High multipoles of CMB B-modes (e.g. $\ell\gtrsim100$) are significantly affected by CMB lensing.
The detection of a B-mode signal would provide valuable information and would help us to test modified gravity theories \cite{Ade:2014xna,Ade:2014afa}.

In modified gravity theory, the evolution equation of gravitational waves is presented in a nonstandard way \cite{Dubovsky:2009xk,Amendola:2014wma,Lin:2016gve,Xu:2014uba,Brax:2017pzt,Pettorino:2014bka,Raveri:2014eea,Boubekeur:2014uaa}. In general, the evolution of the gravitational wave amplitude can be changed by three ways: (1) the damping rate of gravitational waves; (2) the dispersion relation; (3) an additional source term on the perfect fluid.
Any modification of the tensor wave equation can potentially lead to observable effects on the CMB on both temperature and polarization spectrum.
In this paper, we explore the imprint of the modified dispersion relation of graviton in the CMB angular power spectra. Phenomenologically we suppose that the dispersion relation of graviton is parametrized by
\e
E^2=p^2+\sum_n \lambda_n {p^n \over M^{n-2}},
\label{dispm}
\q
where $p$ is the physical momentum of graviton and $M$ is an energy scale. For $n=0$, it implies that the graviton has a mass $\sqrt{\lambda_0}M$. Since the CMB itself reflects the physics at very large scale, the terms with $n>0$ which encode the modification at UV scales should not significantly affect the CMB power spectra, and thus we focus on the cases with $n< 0$ which may imply the non-locality of gravity. Note that $n=0$ has been investigated in \cite{Dubovsky:2009xk}. Here we suppose that the evolution of the background cosmology is still described by the usual Friedmann equation and explore how the evolution of gravitational waves corresponding to the above dispersion relation affect the CMB power spectra. For $n<0$, the modes with small momenta are significantly changed by the modified dispersion relation and acquire time-dependent oscillations which enhance the B-mode spectrum at large scales (or low multipoles).

This paper is organized as follows. In Sec.~2, we provide the detail to illustrate the effects of the modified dispersion relation of graviton on CMB power spectra by semi-analytical and full numerical methods by modifying the publicly available codes CAMB \cite{Hojjati:2011ix,Dossett:2011tn}. A short summary and discussion will be given in Sec.~3.


\section{The tensor contribution to the cosmic microwave background spectra}
In the synchronous gauge, the metric with perturbations is given by
\e
ds^2=a^2(\tau)\{-d\tau^2+(\delta_{ij}+h_{ij})dx^idx^j\}\label{metric},
\q
where the components $g_{00}$ and $g_{0i}$ are unperturbed.
The temperature and polarization perturbations generated by gravitational waves satisfy the following Boltzmann equations, \cite{Ma:1995ey,Zaldarriaga:1996xe,Seljak:1996is,Zaldarriaga:1995gi,Cosmology,ModernCosmology},
\m
&\dot{\tilde{\Delta}}_T^{(T)}+ik\mu\tilde{\Delta}_T^{(T)}=-\dot{h}-\dot{\kappa}[\tilde{\Delta} _T^{(T)}-\Psi], \\
&\dot{\tilde{\Delta}}_P^{(T)}+ik\mu\tilde{\Delta}_P^{(T)}=-\dot{\kappa}[\tilde{\Delta} _P^{(T)}+\Psi], \\
&\Psi\equiv\left[\frac{1}{10}\tilde{\Delta}_{T0}^{(T)}+\frac{1}{7}\tilde{\Delta}_{T2}^{(T)}+\frac{3}{70}\tilde{\Delta}_{T4}^{(T)}-\frac{3}{5}\tilde{\Delta}_{P0}^{(T)}
+\frac{6}{7}\tilde{\Delta}_{P2}^{(T)}-\frac{3}{70}\tilde{\Delta}_{P4}^{(T)}\right]\label{Boltzmann equation},
\n
where the dots denote the derivative with respect to the conformal time $\tau\equiv \int dt/a(t)$, $\mu=\hat{n}\cdot\hat{k}$ is the angle between the photon direction and wave vector, and the superscript $(T)$ denotes the contributions from the tensor perturbations. Here $\dot{\kappa}=an_ex_e\sigma_T$ is the differential optical depth for Thomson scattering, where $a(\tau)$ is the expansion factor normalized to unity today, $n_e$ is the electron density, $x_e$ is the ionization fraction and $\sigma_T$ is the Thomson cross section. Temperature anisotropies have additional source in the metric perturbation $h$. The only external source is that of the tensor metric perturbation which evolves according to the Einstein equations.

These Boltzmann equations can be integrated along the line of sight to give, \cite{Seljak:1996is},
\m
\Delta^{(T)}_{(T,P)\ell}&=&\int_0^{\tau_0} d\tau S^{(T)}_{{T,P}}(k,\tau)\chi^\ell_k(\tau_0-\tau), \\
S_T^{(T)}(k,\tau)&=&-\dot{h}e^{-\kappa}+g\Psi, \\
S_P^{(T)}(k,\tau)&=&-g\Psi,
\n
where $g(\tau)=\dot{\kappa}\exp(-\kappa)$ is the visibility function, $\chi^\ell_k(\tau)=\sqrt{\frac{(\ell+2)!}{2(\ell-2)!}}\frac{j_\ell(k\tau)}{(k\tau)^2}$ is the spherical Bessel function. The anisotropy spectrum or the polarization spectrum is then obtained by integrating over the initial power spectrum of the metric perturbation
\e
C_\ell^{(T)}=(4\pi)^2\int k^2dkP_h(k)\left|\Delta_{(T,P)\ell}^{(T)}(k,\tau=\tau_0)\right|^2\label{power spectrum}.
\q
The expressions for the E and B power spectrum are given by
\m
C_{E\ell}^{(T)}&=&(4\pi)^2\int k^2dk P_h(k)\left[\int_0^{\tau_0}d\tau S^{(T)}_P\left(-j_\ell(x)+j_\ell^{\prime\prime}(x)+\frac{2j_\ell(x)}{x^2}+\frac{4j_\ell^\prime(x)}{x}\right)\right]^2, \\
C_{B\ell}^{(T)}&=&(4\pi)^2\int k^2dk P_h(k)\left[\int_0^{\tau_0}d\tau S^{(T)}_P\left(2j_\ell^\prime(x)+\frac{4j_\ell(x)}{x}\right)\right]^2,
\n
where $x=k(\tau_0-\tau)$. The two-point correlations of the temperature anisotropies and polarization patterns at different points in the sky are presented by these equations.
In the limit of vanishing momentum, only $\tilde{\Delta}_{T,0}^{(T)}$ and $\tilde{\Delta}_{P,0}^{(T)}$ are non-zero and the solutions of Boltzmann equations are given by, \cite{M. M. Basko},
\m
\tilde{\Delta}_{T,0}^{(T)}(\tau_0)&=&\int^{\tau_0}_0d\tau (-\frac{6}{7}e^{-\kappa(\tau)}-\frac{1}{7}e^{-\frac{3}{10}\kappa(\tau)})\dot{h}, \label{sdt}\\
\tilde{\Delta}_{P,0}^{(T)}(\tau_0)&=&\int^{\tau_0}_0d\tau(-\frac{1}{7}e^{-\kappa(\tau)}+\frac{1}{7}e^{-\frac{3}{10}\kappa(\tau)})\dot{h}\label{delta}
\n
which clearly show that the transfer functions $\tilde{\Delta}_{T,0}^{(T)}$ and $\tilde{\Delta}_{P,0}^{(T)}$ are closely related to the time evolution of tensor perturbations.


We mainly focus on the effects of evolution of gravitational waves with modified dispersion relation on the CMB. Since the typical energy scale relevant to CMB is the Hubble scale $H_r\approx 2.3 \times10^4 H_0 \approx 3.3\times 10^{-29}$ eV at recombination, for convenience, the energy scale $M$ in Eq.~(\ref{dispm}) can be taken as $M=H_r$. Here, for example, we only pick up one term in the second part of Eq.~(\ref{dispm}), and then the equation of motion for the gravitational wave becomes
\e
\ddot{h}_k+2\frac{\dot{a}}{a}\dot h_k+\left(k^2+\lambda_\alpha \frac{k^\alpha}{H_r^{\alpha-2}}a^{2-\alpha}\right)h_k=0
\label{waveequation},
\q
where contributions on the right hand side due to the anisotropic stress generated by neutrinos and photons are ignored, and $k$ is the comoving mode. For $\alpha=0$, it nothing but the case for a massive graviton with mass $m_g=\sqrt{\lambda_0} H_r$. Similarly, we introduce a parameter
\e
m_k^2\equiv \lambda_\alpha \(k\over k_r\)^\alpha H_r^2
\q
which is the physical effective mass of graviton with comoving mode $k$ at recombination, where $k_r\equiv a_r H_r$ is the mode which re-entered horizon at recombination and $a_r$ is the scale factor. Since CMB measures the structures at cosmic scales and then is a powerful tool to probe gravity in the infrared (IR) limit, we focus on the case of $\alpha\leq 0$ in this paper.
The short-wavelength modes satisfying
\m
m_k\ll {k\over a_r},
\n
or equivalently
\m
k\gg k_0\equiv \lambda_\alpha^{\frac{1}{2-\alpha}}k_r,
\n
are not affected by modified term in the dispersion relation, no matter $m_k$ is below the Hubble rate $H_r$ or not. In terms of the multipole number, the transition to the massless regime corresponds to
\m
\ell_0\equiv k_0(\tau_0-\tau_r)\approx {k_r\over H_0} \int^{z_r}_0 {dz\over E(z)}
\approx 66.2\times \lambda_\alpha^{1\over 2-\alpha},
\n
where $E(z)\equiv H(z)/H_0=\sqrt{\Omega_\Lambda+\Omega_m (1+z)^3+\Omega_r (1+z)^4}$.
Therefore the CMB power spectra are expected to be the same as those in the massless case for $\ell\gg \ell_0$.
If $\lambda_\alpha<1$, there is a parameter space for
\m
\frac{k}{a_r}< m_k<H_r\label{condition2},
\n
or equivalently
\m
\lambda_\alpha^{-\frac{1}{\alpha}}k_r<k<\lambda_\alpha^{\frac{1}{2-\alpha}}k_r.
\n
Even though the second term in the bracket of Eq.~(\ref{waveequation}) becomes dominant, these modes do not oscillate because of the Hubble friction term. Finally, the modes start to oscillate before the recombination if their wavelengths are long enough, namely
\m
m_k>\frac{k}{a_r}\quad \hbox{and} \quad m_k>H_r,
\n
which implies
\m
k<\hbox{min}\(\lambda_\alpha^{-\frac{1}{\alpha}},\ \lambda_\alpha^{\frac{1}{2-\alpha}}\) k_r.
\n
For $\alpha=0$, this condition becomes $k<\lambda_0^{1/2}k_r$ and $\lambda_0>1$.

\begin{figure}[H]
\centering
\includegraphics[width=12cm]{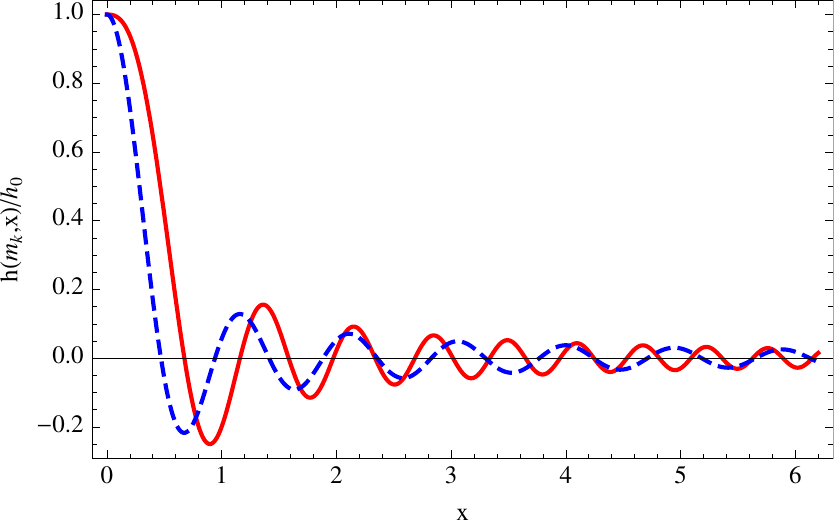}\\
\includegraphics[width=12cm]{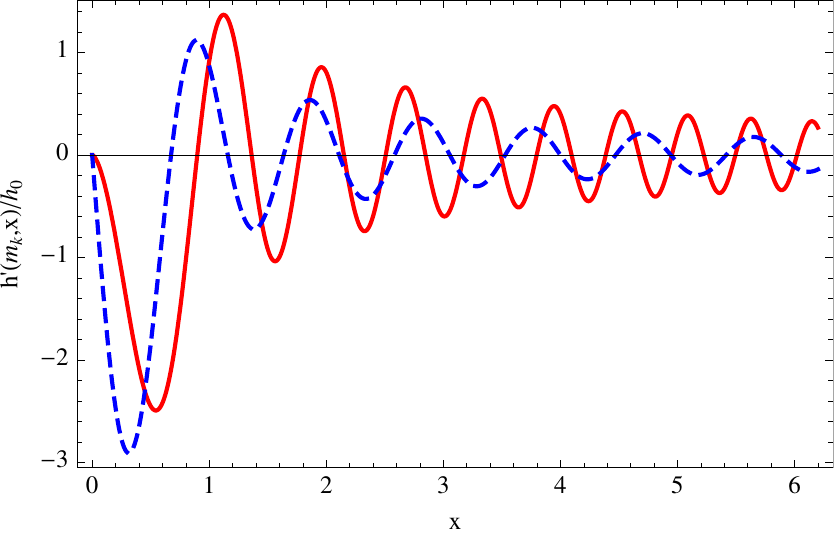}
\caption{The plots of ${h(m_k,x)/ h_0}$ and ${h'(m_k,x)/ h_0}$ for $\alpha=-1$ (red solid curves) and $\alpha=0$ (blue dashed curves) respectively, where $m_k/H_r=10$ is kept fixed.
}
\label{fig:hhp}
\end{figure}

At the matter dominated stage, $a(t)=a_r (t/t_r)^{2/3}$, and for the long wavelength modes satisfying $k\ll \hbox{min}\(\lambda_\alpha^{-\frac{1}{\alpha}},\ \lambda_\alpha^{\frac{1}{2-\alpha}}\) k_r$, Eq.~(\ref{waveequation}) can be simplified to be
\m
h_k''+{2\over x}h_k'+(m_k t_r)^2 x^{-2\alpha/3}h_k=0,
\label{longwave}
\n
where the prime denotes the derivative with respect to $x\equiv t/t_r$. Note $m_k t_r={2m_k\over 3H_r}={2\over 3} \lambda_\alpha^{1/2} \({k\over k_r}\)^{\alpha/2}$ because of $H_r=2/(3t_r)$. Eq.~(\ref{longwave}) can be analytically solved and the solution is given by
\m
{h(m_k,x)\over h_0}= 2^{\beta} \Gamma\(1+\beta\) {\bar x}^{-\beta} J_\beta({\bar x})
\label{h},
\n
and then
\m
{h'(m_k,x)\over h_0}= -2^\beta \Gamma\(1+\beta\) m_k t_r x^{{1\over 2\beta}-1} {\bar x}^{-\beta} J_{1+\beta}({\bar x}),
\n
where
\m
\beta&=&\frac{3}{6-2\alpha},\\
{\bar x}&=&2\beta m_k t_r x^{\frac{1}{2\beta}},
\n
and $J_\beta({\bar x})$ is the Bessel function of the first kind. For $\alpha=0$, ${h(m_k,x)/h_0}=\sin(m_g t)/(m_g t)$ which is that same as that in \cite{Dubovsky:2009xk}.
The behaviors of ${h(m_k,x)/ h_0}$ and ${h'(m_k,x)/ h_0}$ for $\alpha=-1$ and $\alpha=0$ are illustrated in Fig.~\ref{fig:hhp}.

\begin{figure}[h]
\centering
\includegraphics[width=12cm]{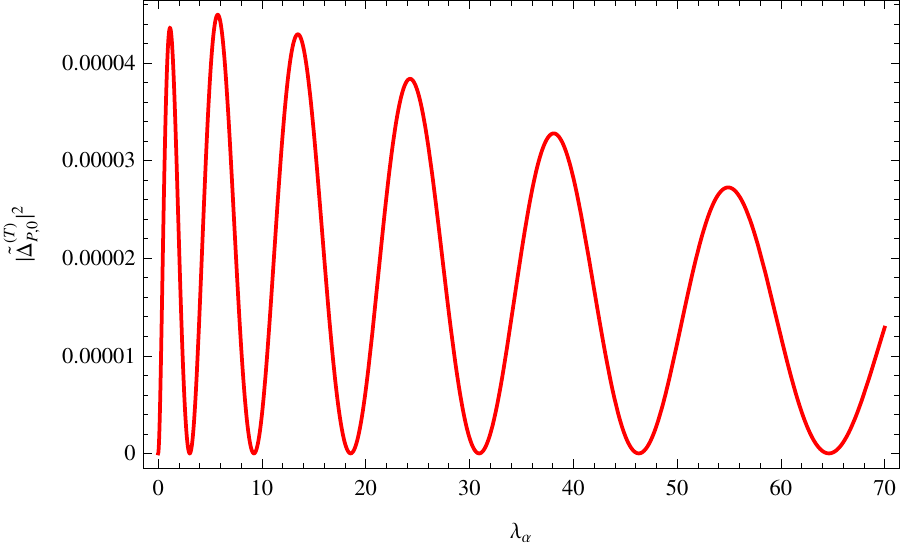}
\caption{The plot of $\left|\tilde\Delta_{P,0}^{(T)}(\tau_0)\right|^2$ for $k=0.035k_r$, where $\alpha=-1$.}
\label{fig:deltap}
\end{figure}

\begin{figure}[h]
\centering
\includegraphics[width=12cm]{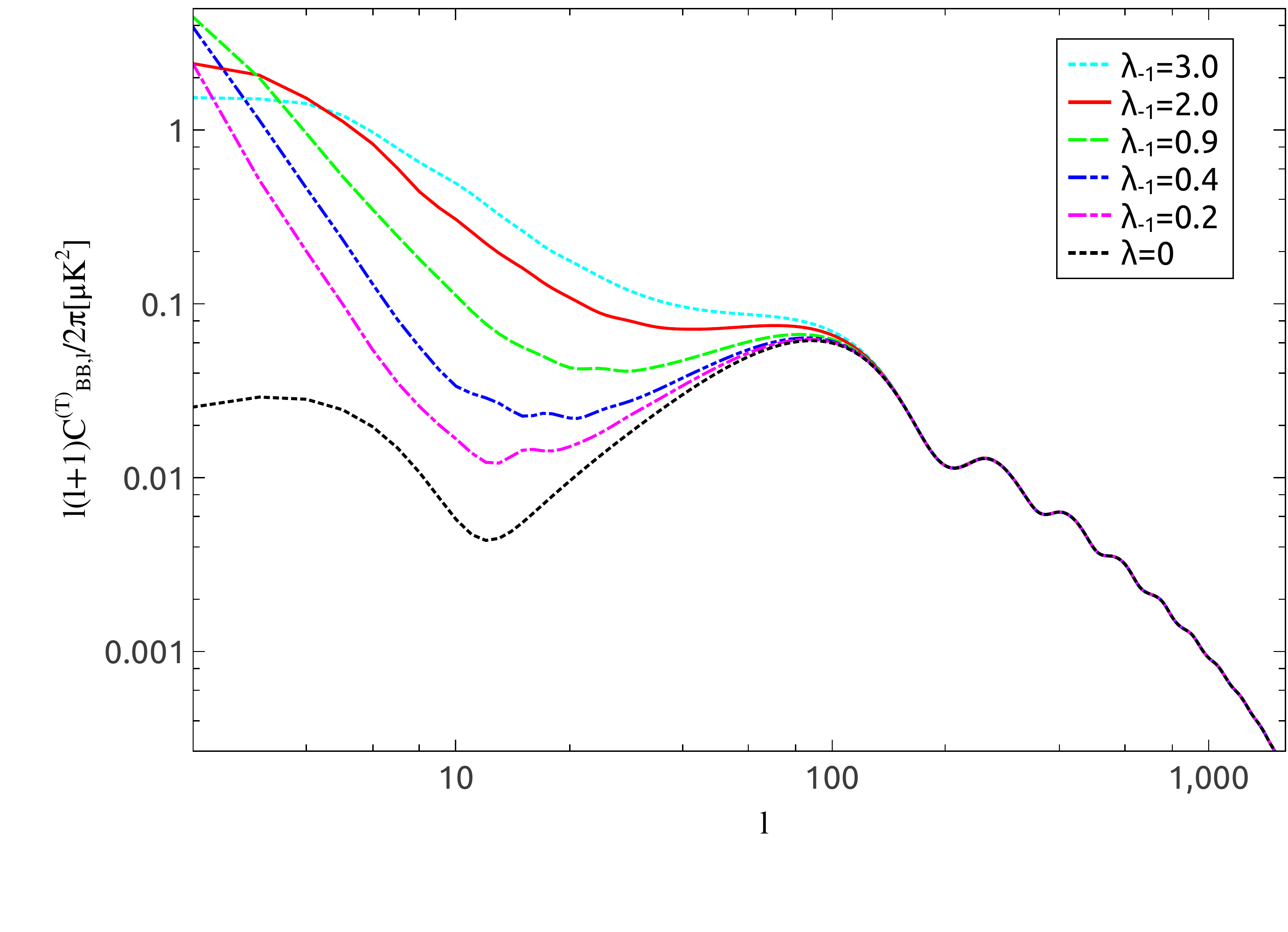}\\
\includegraphics[width=12cm]{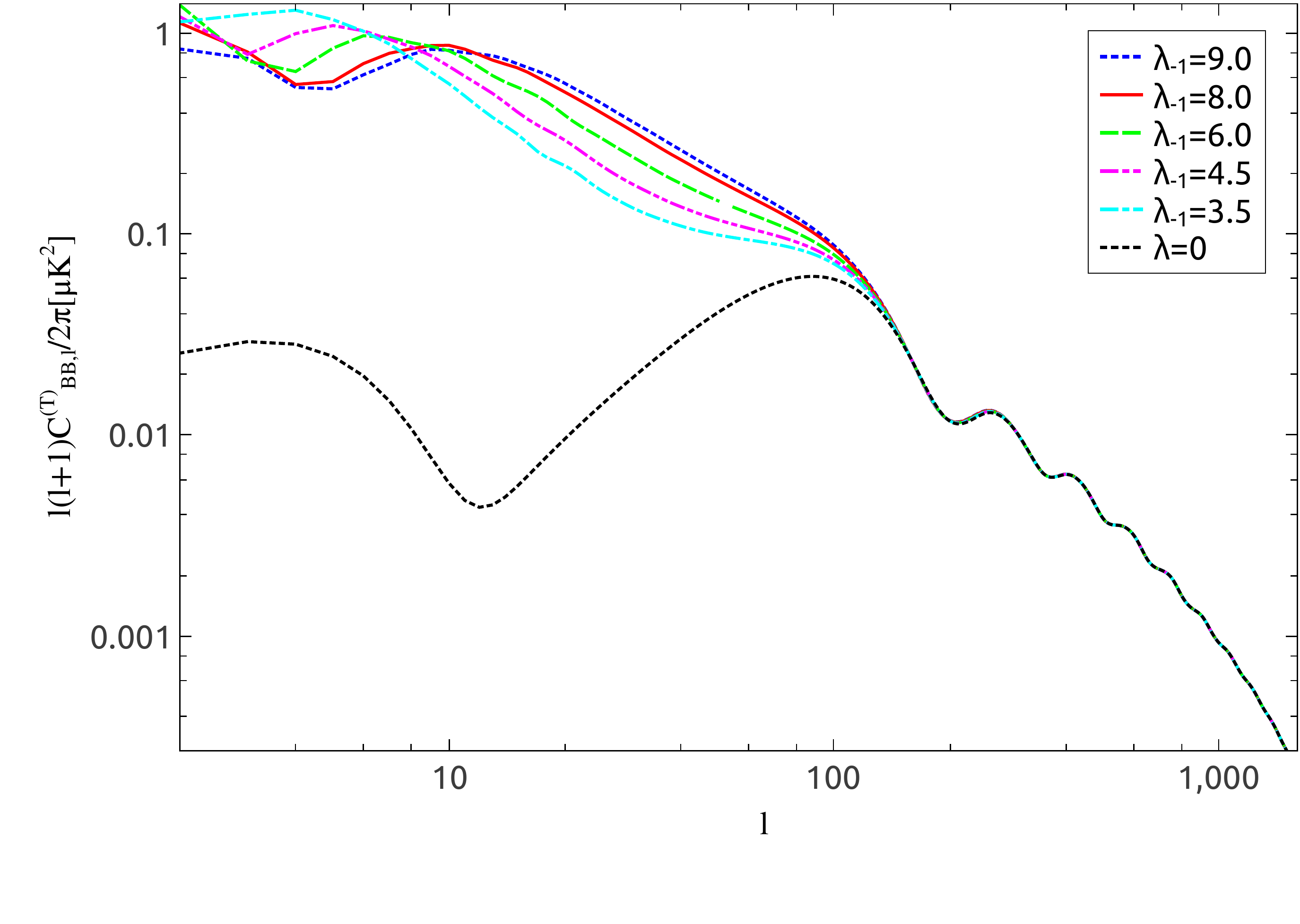}
\caption{The plots of $C^{(T)}_{BB,\ell}$ for different values of $\lambda_{\alpha=-1}$. The black dotted curve corresponds to the massless case.}
\label{fig:BB}
\end{figure}

In the vanishing momentum limit, the non-zero solutions of Boltzmann equations are given in Eqs.~(\ref{sdt}) and (\ref{delta}) which indicates that the oscillatory solution of gravitational wave in Eq.~(\ref{h}) will enhance CMB angular power spectra at large scales. Let us consider a mode with $k\ll k_r$, e.g. $k=0.035k_r$ which roughly corresponds to the CMB quadrupole. Integrating over the conformal time in Eq.~(\ref{delta}), the function of $\left|\tilde\Delta_{P,0}^{(T)}(\tau_0)\right|^2$ is illustrated in Fig.~\ref{fig:deltap} for $\alpha=-1$.
The oscillation signature comes from the gravitational waves and the envelope gradually decreases with $\lambda_{-1}$ owning to the damping term in the wave equation.
Roughly speaking, $\left|\tilde\Delta_{P,0}^{(T)}(\tau_0)\right|^2$ monotonously increases  from $\lambda_{-1}=0$ to $1$, and then drops from 1 to 3.
Because of
\m
C^{(T)}_{BB,2}\propto\left|\tilde\Delta_{P,0}^{(T)}(\tau_0)\right|^2,
\n
the oscillatory behavior in Fig.~\ref{fig:deltap} is expected to yield oscillatory $C^{(T)}_{BB,2}$ for different values of $\lambda_{-1}$. These signatures match the numerical results for BB tensor angular power spectrum at $\ell=2$ quite well in Fig.~\ref{fig:BB}.
In addition, for $0<\lambda_{-1}\lesssim 3$, $C^{(T)}_{BB,\ell}$ is oscillatory for $\ell\lesssim 5$, but keeps increasing for $5\lesssim \ell \lesssim 100$. See the upper plot in Fig.~\ref{fig:BB}. For $3\lesssim \lambda_{-1}\lesssim 9$, $C^{(T)}_{BB,\ell=2}$ is also oscillatory, but keeps decreasing for $3\lesssim \ell \lesssim 8$ and increasing for $8\lesssim \ell \lesssim 100$. See the bottom plot in Fig.~\ref{fig:BB}.

\begin{figure}
\centering
\includegraphics[width=18cm]{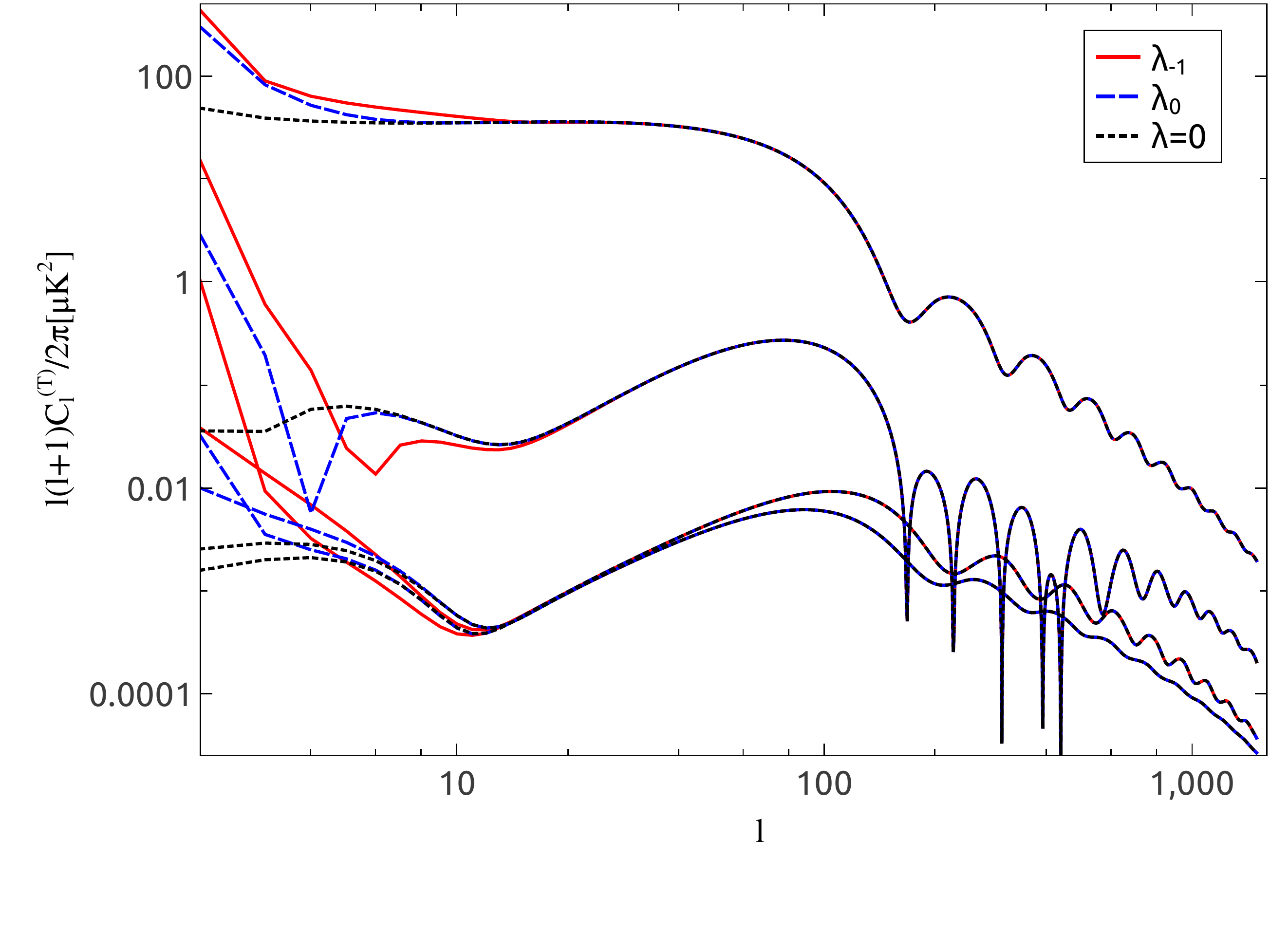}
\caption{The plots of TT, TE, EE and BB angular power spectra contributed from tensor perturbations. Here we take $r=0.1$ and compare the results for $\alpha=0$ (blue dashed curves) and $\alpha=-1$ (red solid curves) with $\lambda_\alpha=10^{-6}$. The black dotted curves correspond to $\lambda_\alpha=0$. }
\label{fig:tensor}
\end{figure}

In order to obtain the numerical results for the CMB angular power spectra for the modified dispersion relation of graviton, we modify CAMB \cite{Hojjati:2011ix} by taking into account the evolution equation for the tensor perturbation in Eq.~(\ref{waveequation}), where we add two parameters $\alpha$ and $\lambda_\alpha$. By varying these two parameters, we can explore how the modified dispersion relation of graviton affects the CMB power spectra. 
Keeping $\lambda_\alpha=10^{-6}$, we plot TT, TE, EE and BB angular power spectra contributed from the tensor perturbations for $\alpha=-1$ and $\alpha=0$ in Fig.~\ref{fig:tensor}.
We find that the CMB polarization power spectra are more sensitive to the modification of the dispersion relation of graviton. Keeping $m_k$ fixed, the oscillation of gravitational wave for $\alpha=-1$ is slightly stronger than $\alpha=0$ in Fig.~\ref{fig:hhp}, and thus all of the quadrupoles of TT, TE, EE and BB power spectra for $\alpha=-1$ is higher than those for $\alpha=0$.

\section{Summary and discussion}
\label{sd}

In this paper we explore how the modified dispersion relation of graviton affects the shape of CMB angular power spectra. Due to the oscillations of the amplitude of gravitational waves, the modification of dispersion relation enhances the quadrupoles of CMB power spectra, and may lead to different oscillatory behaviors for different multipoles with $\ell\lesssim 100$.

Once the CMB B-modes contributed by primordial gravitational waves are detected in the future, we can test the dispersion relation of graviton at the energy scales around $10^{-29}$ eV. In particular, the low CMB multipoles $(\ell \lesssim 10)$ are very sensitive to the modification of dispersion relation of graviton, and it implies that the low CMB multipoles are very useful to test the dispersion relation of graviton.


\vspace{5mm}
\noindent {\bf Acknowledgments}

We acknowledge the use of HPC Cluster of ITP-CAS.
This work is supported by grants from NSFC (grant NO. 11335012, 11575271, 11690021), Top-Notch Young Talents Program of China, and partly supported by Key Research Program of Frontier Sciences, CAS.




\begin{thebibliography}{99}
\frenchspacing

\bibitem{Clifton:2011jh}
  T.~Clifton, P.~G.~Ferreira, A.~Padilla and C.~Skordis,
  Phys.\ Rept.\  {\bf 513}, 1 (2012)
  doi:10.1016/j.physrep.2012.01.001
  [arXiv:1106.2476 [astro-ph.CO]].


\bibitem{Nojiri:2010wj}
  S.~Nojiri and S.~D.~Odintsov,
  Phys.\ Rept.\  {\bf 505}, 59 (2011)
  doi:10.1016/j.physrep.2011.04.001
  [arXiv:1011.0544 [gr-qc]].


\bibitem{Capozziello:2011et}
  S.~Capozziello and M.~De Laurentis,
  Phys.\ Rept.\  {\bf 509}, 167 (2011)
  doi:10.1016/j.physrep.2011.09.003
  [arXiv:1108.6266 [gr-qc]].




\bibitem{Kamionkowski:1996zd}
  M.~Kamionkowski, A.~Kosowsky and A.~Stebbins,
  Phys.\ Rev.\ Lett.\  {\bf 78}, 2058 (1997)
  doi:10.1103/PhysRevLett.78.2058
  [astro-ph/9609132].


\bibitem{Cabella:2004mk}
  P.~Cabella and M.~Kamionkowski,
  [astro-ph/0403392].





\bibitem{Kosowsky:1998mb}
  A.~Kosowsky,
  New Astron.\ Rev.\  {\bf 43}, 157 (1999)
  doi:10.1016/S1387-6473(99)00009-3
  [astro-ph/9904102].

\bibitem{Kamionkowski:2015yta}
  M.~Kamionkowski and E.~D.~Kovetz,
  Ann.\ Rev.\ Astron.\ Astrophys.\  {\bf 54}, 227 (2016)
  doi:10.1146/annurev-astro-081915-023433
  [arXiv:1510.06042 [astro-ph.CO]].


\bibitem{Radiationtransfer}
S. Chandrasekhar, Radiation Transfer (Dover, 1960).



\bibitem{Ade:2014xna}
  P.~A.~R.~Ade {\it et al.} [BICEP2 Collaboration],
  Phys.\ Rev.\ Lett.\  {\bf 112}, no. 24, 241101 (2014)
  doi:10.1103/PhysRevLett.112.241101
  [arXiv:1403.3985 [astro-ph.CO]].

\bibitem{Ade:2014afa}
  P.~A.~R.~Ade {\it et al.} [POLARBEAR Collaboration],
  Astrophys.\ J.\  {\bf 794}, no. 2, 171 (2014)
  doi:10.1088/0004-637X/794/2/171
  [arXiv:1403.2369 [astro-ph.CO]].


\bibitem{Dubovsky:2009xk}
  S.~Dubovsky, R.~Flauger, A.~Starobinsky and I.~Tkachev,
  Phys.\ Rev.\ D {\bf 81}, 023523 (2010)
  [arXiv:0907.1658 [astro-ph.CO]].


\bibitem{Lin:2016gve}
  W.~Lin and M.~Ishak,
  Phys.\ Rev.\ D {\bf 94}, no. 12, 123011 (2016)
  [arXiv:1605.03504 [astro-ph.CO]].

\bibitem{Brax:2017pzt}
  P.~Brax, S.~Cespedes and A.~C.~Davis,
  [arXiv:1710.09818 [astro-ph.CO]].


\bibitem{Amendola:2014wma}
  L.~Amendola, G.~Ballesteros and V.~Pettorino,
  Phys.\ Rev.\ D {\bf 90}, 043009 (2014)
  [arXiv:1405.7004 [astro-ph.CO]].


\bibitem{Xu:2014uba}
  L.~Xu,
  Phys.\ Rev.\ D {\bf 91}, 103520 (2015)
  [arXiv:1410.6977 [astro-ph.CO]].

\bibitem{Pettorino:2014bka}
  V.~Pettorino and L.~Amendola,
  Phys.\ Lett.\ B {\bf 742}, 353 (2015)
  doi:10.1016/j.physletb.2015.02.007
  [arXiv:1408.2224 [astro-ph.CO]].

\bibitem{Raveri:2014eea}
  M.~Raveri, C.~Baccigalupi, A.~Silvestri and S.~Y.~Zhou,
  Phys.\ Rev.\ D {\bf 91}, no. 6, 061501 (2015)
  doi:10.1103/PhysRevD.91.061501
  [arXiv:1405.7974 [astro-ph.CO]].



\bibitem{Boubekeur:2014uaa}
  L.~Boubekeur, E.~Giusarma, O.~Mena and H.~Ram¨ªrez,
  Phys.\ Rev.\ D {\bf 90}, no. 10, 103512 (2014)
  doi:10.1103/PhysRevD.90.103512
  [arXiv:1407.6837 [astro-ph.CO]].



\bibitem{Hojjati:2011ix}
  A.~Hojjati, L.~Pogosian and G.~B.~Zhao,
  JCAP {\bf 1108}, 005 (2011)
  doi:10.1088/1475-7516/2011/08/005
  [arXiv:1106.4543 [astro-ph.CO]].

\bibitem{Dossett:2011tn}
  J.~N.~Dossett, M.~Ishak and J.~Moldenhauer,
  Phys.\ Rev.\ D {\bf 84}, 123001 (2011)
  doi:10.1103/PhysRevD.84.123001
  [arXiv:1109.4583 [astro-ph.CO]].

\bibitem{Ma:1995ey}
  C.~P.~Ma and E.~Bertschinger,
  Astrophys.\ J.\  {\bf 455}, 7 (1995)
  doi:10.1086/176550
  [astro-ph/9506072].



\bibitem{Zaldarriaga:1996xe}
  M.~Zaldarriaga and U.~Seljak,
  Phys.\ Rev.\ D {\bf 55}, 1830 (1997)
  [astro-ph/9609170].


\bibitem{Seljak:1996is}
  U.~Seljak and M.~Zaldarriaga,
  Astrophys.\ J.\  {\bf 469}, 437 (1996)
  [astro-ph/9603033].


\bibitem{Zaldarriaga:1995gi}
  M.~Zaldarriaga and D.~D.~Harari,
  Phys.\ Rev.\ D {\bf 52}, 3276 (1995)
  [astro-ph/9504085].

\bibitem{Cosmology}
Steven Weinberg, Cosmology (Oxford University Press,
New York, 2008).

\bibitem{ModernCosmology}
Scott Dodelson, Modern Cosmology (Academic Press,
San Diego, 2003).


\bibitem{M. M. Basko}
  M.~M.~Basko and A.~G.~Polnarev,
  Mon.\ Not.\ R.\ astr.\ Soc.\ {\bf 191}, 207 (1980).






\end{thebibliography}
\end{document}